\documentclass[11pt,graphicx,amsmath]{article}
\usepackage{amsmath}
\usepackage{graphicx}
\usepackage{bm}

\title{ Statefinder parameters for
quantum effective  Yang-Mills condensate dark energy model }

\author{\small   Minglei  Tong\thanks{mltong@mail.ustc.edu.cn}\ ,
                 Yang Zhang and  Tianyang  Xia \\
        \small Astrophysics Center \\
       \small University of Science and Technology of China \\
       \small Hefei, Anhui, China }
 \date{}

\begin{document}
\maketitle

\def\be{\begin{equation}}
\def\ee{\end{equation}}
\def\ba{\begin{eqnarray}}
\def\ea{\end{eqnarray}}
\def\nn{\nonumber}

 \baselineskip=19truept


\begin{center}
\Large  Abstract
\end{center}

\begin{quote}
\Large

The quantum effective  Yang-Mills condensate (YMC) dark energy model
has some distinguished features that it naturally solves the
coincidence problem and, at the same time, is able to give an
equation of state $w$ crossing $-1$. In this work we further employ
the Statefinder pair $(r,s)$ introduced by Sahni et al to diagnose
the YMC model for three cases: the non-coupling, the  YMC decaying
into  matter only, and the YMC decaying into both  matter and
radiation. The trajectories  $(r,s)$ and  $(r,q)$, and the
evolutions $r(z)$, $s(z)$ are explicitly presented. It is found
that, the YMC model in all three cases has  $r\simeq 1$ for  $ z <
10$ and $s\simeq 0$ for $z<5$ with only small deviations  $\simeq
0.02$, quite close to the cosmological constant model (LCDM), but is
obviously differentiated from other dark energy models, such as
quiesence, kinessence etc.

\end{quote}

{\bf keywords}: Yang-Mills condensate, dark energy, Statefinder


\newpage

\large
 \baselineskip=18truept

\section{ Introduction }

The currently accelerating expansion of the universe
has been indicated by the observation
of Type Ia supernovae   \cite{Riess},
and is consistent with the data from
cosmic microwave background (CMB)  \cite{Bernardis,Bennett} and
from the cosmic large scale structure \cite{Bahcall}.
The acceleration of the expanding universe
is attributed to the
dominant dark energy $\Omega_\Lambda \simeq 0.73$,
coexisting with the matter $\Omega_m \simeq 0.27$.
The simplest model for dark energy is
the cosmological constant \cite{Weinberg},
with an equation of state (EOS)  $w=-1$
as the universe evolves.
However,  two questions arise  from this scenarios,
namely, the fine-tuning problem and the cosmic coincidence problem.
While the former exists for almost all the
 dark energy models,
the latter has been addressed in the class of dynamical dark energy
models, which take some dynamically-evolving field as the candidate
for the dark energy. Among them are the quintessence
\cite{Wetterich}, phantom \cite{Caldwell}, k-essence
\cite{Armendariz}, quintom \cite{Wei}, tachyonic \cite{Padmanabhan},
holographic dark energy model \cite{Li}, and interacting dark energy
model \cite{Amendola}. Besides, there is another interesting type of
model built on the quantum effective gravity \cite{Parker}, which
includes the quantum corrections of gravitational field to Einstein
equations. Different from these models, the Yang-Mills condensate
dark energy model is based on a vector-type of the  quantum
effective Yang-Mills field
 \cite{Zhang1,zhang0304,Zhang,Zhao1,Xia}.
From field-theoretical point of view, the model has the following
interesting properties: the gauge fields are indispensable to the
Standard Model of particle physics, and the effective Lagrangian of
YMC is determined from the standard  field-theoretical calculations
for each order of loops of quantum corrections, and thereby gives
the correct trace-anomaly, and there is room for change its form by
hand. Moreover, it is found that, for quite generic initial
conditions, the YMC dark energy model always has the desired
tracking behavior that naturally solves the coincidence problem.
This has been accomplished for the cases of 1-loop
\cite{Zhang,Zhao1}, 2-loop  \cite{Xia}, and 3-loop \cite{Wang}
quantum corrections, either with coupling or without coupling to
matter and/or radiation. When coupling with  matter, or with both
matter and radiation, the YMC has  an EOS crossing $-1$ smoothly and
taking  $w\sim-1.05$, as indicated by the recent preliminary
observational data of  the Supernova Legacy Survey
\cite{Astier,Alam1}.

In order to differentiate these various dark energy models,
Sahni et al \cite{Sahni} introduce
a new geometrical diagnostic pair $(r,s)$, called Statefinder,
which involves the third order time-derivative of scale factor.
The pair is related to the EOS of dark energy and its
time derivative.
From the observational side,
the values of $(r,s)$
can be extracted from data coming
form SNAP type experiments \cite{Albert}.
The Statefinder diagnosis  has been applied
to several dark energy models  \cite{Alam,Gorini,Cai,Zhangjingfei}.
In particular, the spatially flat  LCDM
has a fixed point $(r,s)=(1,0)$.
For the 1-loop YMC dark energy model,
Ref.\cite{Zhao} studies the  non-coupling case
with  the radiation contribution being neglected.
In this paper
for a complete treatment of the Statefinder diagnosis,
we work with the 2-loop coupling YMC model
and include the radiation component.
Thereby, there arise
considerable modifications to the 1-loop non-coupling case.

\section{ 2-loop YMC  dark energy model}

We consider a
spatially flat($k=0$) Robertson-Walker (RW) universe,
whose  expansion is determined by the Friedmann equations
\ba  \label{fri1}
&&H^2=\frac{8\pi G}{3}\rho,\\
&&\frac{\ddot{a}}{a}=-\frac{4\pi G}{3}(\rho+3p),\label{fri2}
\ea
where $H=\frac{\dot{a}}{a}$, the pressure $p(t)=p_y+p_r$,
the  energy density
$\rho(t)=\rho_y+\rho_m+\rho_r$,
with the subscripts `$y$', `$m$' and `$r$' refer to
the  YMC dark energy,
the matter (including both baryons and dark matter),
and the radiation, respectively.
Up to the 2-loop quantum corrections,
the energy density
$\rho_y$ and the pressure $p_y$ of the YMC
are given by \cite{Xia,Wang}
\ba  \label{rho}
&&\rho_y=\frac{b}{2}F\left[y+1+\eta\left(\ln|y-1+\delta|
         +\frac{2}{y-1+\delta}\right)\right],\\
&&p_y=\frac{b}{6}F\left[y-3+\eta\left(\ln|y-1+\delta|
         -\frac{2}{y-1+\delta}\right)\right] \label{p},
\ea
where $y\equiv\ln|F/\kappa^2|$,
$F \equiv  E^2-B^2 $,
$\kappa$ is the renormalization scale with dimension of squared mass,
$b=\frac{11N}{3(4\pi)^2}$ for the gauge group $SU(N)$ without
fermions,
$\eta\equiv\frac{2b_1}{b^2}\simeq0.84$ with $b_1=\frac{17N^2}{3(4\pi)^4}$
representing the 2-loop contribution,
and the dimensionless constant
$\delta$ is a parameter representing higher order corrections.
For simplicity,  we take the gauge group to be $SU(2)$
and only consider the  `electric' condensate, i.e., $F=E^2$.
The EOS for the YMC  is
\be  \label{eos}
w=\frac{p_y}{\rho_y}=
  \frac{y-3+\eta(\ln{|y-1+\delta|}-\frac{2}{y-1+\delta})}
   {3[y+1+\eta(\ln{|y-1+\delta|}+\frac{2}{y-1+\delta})]}.
\ee
When one sets $\eta=0$ in the above expressions
of Eqs.(\ref{rho}), (\ref{p}), and (\ref{eos}),
the 1-loop model is recovered \cite{zhang0304,Zhang,Zhao1}.
At high energies $y\rightarrow \infty$, $p_y$ is positive,
and the EOS of YMC approaches to that of a radiation
$w\rightarrow1/3$,
as is expected for an effective quantum field theory.

The dynamical evolutions of the three components of the universe are
\ba\label{rhoy}
&&\dot{\rho}_y+3\frac{\dot{a}}{a}(\rho_y+p_y)
           =-\Gamma\rho_y-\Gamma'\rho_y,
\\ \label{rhom}
&&\dot{\rho}_m+3\frac{\dot{a}}{a}\rho_m
           =\Gamma\rho_y,\\
&&\dot{\rho}_r+3\frac{\dot{a}}{a}(\rho_r+p_r)
           =\Gamma'\rho_y, \label{rhor}
\ea
where  $\Gamma$ and $\Gamma'$ are
the decay rate of  YMC
into   matter and   radiation, respectively.
Taking $\rho_y$ in Eq.(\ref{rho}) to be
the dark energy with  $\rho_y(t_0)\sim0.73\rho_c$,
the scale $\kappa$ is fixed by
$\kappa^{1/2}\simeq 7.6 h_0^{1/2}\times10^{-3}$ eV,
where $h_0\sim0.72$ is the current value of the Hubble parameter.
For concreteness, $\delta=3$ and $\eta=0.84$ are taken.

\begin{figure}
\centerline{\includegraphics[width=8cm]{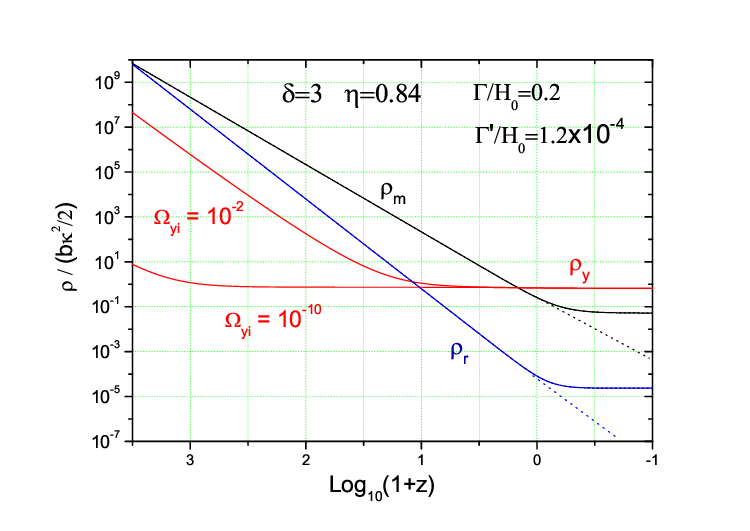}}
 \caption{\label{densityboth}\small
The evolution of the energy densities.
In the non-coupling case $\rho_m \propto a(t)^{-3}$ and $\rho_r \propto a(t)^{-4}$
throughout.
In the case of YMC decaying into matter,
$\rho_m$ deviates from  $  \propto a(t)^{-3}$ around $z\sim 0$
and levels off.
In the case of YMC decaying into both  matter and radiation,
both $\rho_m(t)$ and $\rho_r(t)$ level off  around $z\sim 0$.
Note that two different initial condition: $\Omega_{yi}=10^{-2}$ and $10^{-10}$,
yield the same $\rho_y$, $\rho_m$, and $\rho_r$ at $z\sim 0$. }
\end{figure}

The initial  matter and radiation densities
at $z\simeq 3454$ \cite{Bennett} are taken to be
$x_i=R_i \simeq 1 \times10^{10}$,
where $x\equiv\rho_m/\frac{1}{2}b\kappa^2$,
and $R\equiv\rho_r/\frac{1}{2}b\kappa^2$.
The initial YMC can be chosen to be in a broad range
$  y_i=(1, \,\, 15)$,
corresponding to $ \Omega_{yi} \simeq (10^{-10}, 10^{-2})$.
For the case of YMC decaying into matter,
we take $\Gamma/H_0 =0.2$ with
$H_0 = (\frac{4\pi Gb\kappa^2}{3})^{1/2}$,
approximately the Hubble constant.
For the case of YMC decaying into both matter and radiation,
we take $\Gamma/H_0 = 0.2$ and  $\Gamma'/H_0 =  1.2\times10^{-4}$.
Fig.\ref{densityboth} shows
the evolution of the energy densities for various components
in all three cases.
It is seen that, for the initial values of  $\Omega_{yi}$
ranging eight orders of magnitude,
the present status
$\Omega_y\sim 0.73$,
$\Omega_m \sim  0.27$, and $\Omega_r  \sim 10^{-5}$,
are achieved.
In this sense, the coincidence problem of dark energy
is naturally solved in the YMC model \cite{zhang0304,Zhao1,Xia}.
It is also found that in the coupling cases
  $w$ can be cross $-1$ smoothly \cite{Xia}.

\section{Statefinder }

The Statefinder pair  ($r, s$) are defined  as the following
\cite{Sahni} \be  \label{RS}
r\equiv\frac{\stackrel{\dots}{a}}{aH^3}, \,\,\,\, s \equiv
\frac{r-1}{3(q-1/2)}~, \ee where the deceleration parameter \be
\label{q} q =-\frac{\ddot{a}}{aH^2}=\frac{1}{2}(1+3\Omega_y
w+\Omega_r), \ee with  $\Omega_y=\rho_y/\rho$ and
$\Omega_r=\rho_r/\rho$. For completeness we include $\Omega_r$,
which is important in the early times. Taking time derivative of
Eq.(\ref{fri2}) and making use of Eqs.(\ref{rhoy}),  (\ref{rhom}),
and (\ref{rhor}) yield the following general form of the Statefinder
pair \ba\label{rboth} &&r=1+\frac{9}{2}\Omega_y  w(1+w)
  -\frac{3}{2}\Omega_y \frac{\dot w}{H} +2\Omega_r
  +  \frac{3\Gamma}{2H}\Omega_y w
  +\frac{\Gamma'}{2H}\Omega_y(3  w -1),  \label{R} \\ \label{sboth}
&&s=\frac{3\Omega_y  w (1+ w)-\Omega_y\frac{\dot w}{H}
    +\frac{4}{3}\Omega_r
    +\frac{\Gamma}{H}\Omega_y w
    +\frac{\Gamma'}{3H}\Omega_y(3 w-1)
     }{3\Omega_y w +\Omega_r} \label{S},
\ea which hold actually for a generic dark energy model.

Let us apply the pair to some simple models.
For any LCDM with a non-zero cosmological constant $\Lambda$,
one simply has $(r,s)=(1,0)$,
and  $q(t)=(-1, \frac{1}{2})$.
For the Steady State Universe (SSU) model \cite{Bondi}
with $a(t)=e^{H_0t}$ one also has $(r,s)=(1,0)$,
but it has a fixed $q(t)=-1$.
For the standard cold dark matter model (SCDM),
our calculation gives $(r,s)=(1,\frac{4}{3})$,
which is different from the  value $(r,s)=(1,1)$
often quoted in literatures \cite{Sahni}.
The details of our derivation is given in Appendix.

In the following we present $(r,s)$ in
the YMC model for three cases:
the non-coupling,
the YMC decaying into matter,
the YMC decaying into both matter and radiation.
The  following plots are based on Eqs.(\ref{q}), (\ref{rboth}) and (\ref{sboth}).

\begin{figure}
\centerline{\includegraphics[width=10cm]{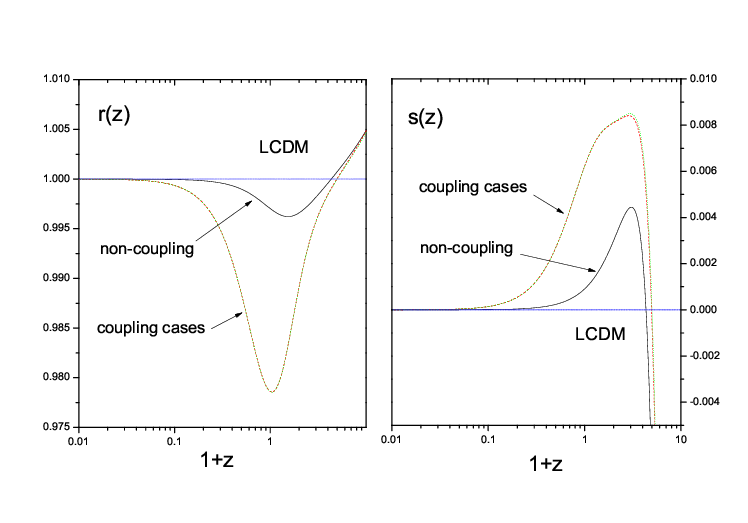}}
 \caption{\label{sztime}
 The left  panel:
 $ r(z) \simeq 1$ for $z<10$ in the three cases of YMC.
 The right panel: $s(z)\simeq 0$ for $z<5$.
 Thus  $(r,s)\simeq (1,0)$ with an error $\simeq  0.02$ in YMC,
  quite  close to that of LCDM.
  The initial YMC fractional energy $\Omega_{yi}=10^{-2}$ is taken.}
 \end{figure}

 \begin{figure}
\centerline{\includegraphics[width=10cm]{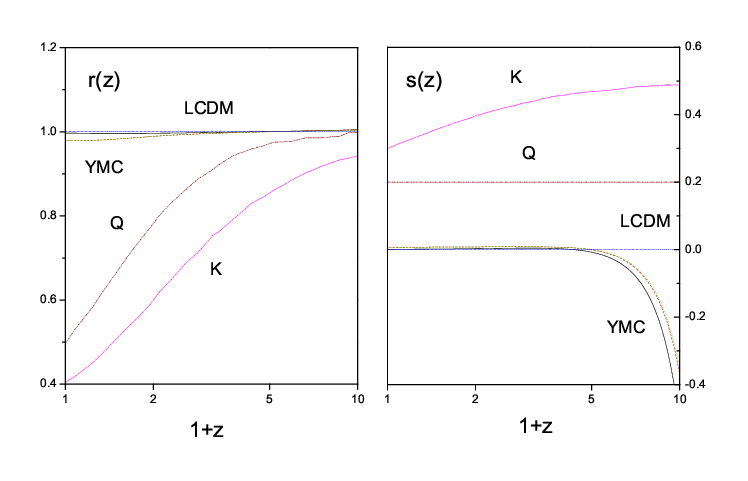}}
 \caption{\label{comp}
  The details of  $r(z)$ on the left
  and $s(z)$ on the right around $z= ( 0, 10)$ in YCM
compared with quiessence (Q) and kinessence (K) models \cite{Sahni}.
 }
 \end{figure}

First, we discuss the evolutions of ($r, s$).
Fig.\ref{sztime} gives  the behaviors of $r(z)$
and $s(z)$ in the recent past
time with $z <  10$ in all the three cases of YMC model.
It is clear that $ r(z) \simeq 1$ for $z<10$
and $s(z)\simeq 0$ for $z<5$.
Thus  $(r,s)\simeq (1,0)$ with an error $\simeq  0.02$ in YMC,
  quite  close to that of LCDM.
The two coupling cases, i.e.,
YMC decaying into matter,
and into both matter and radiation,
yield almost overlapping results
since the YMC-radiation coupling
$\Gamma'/H_0\sim 10^{-4}$ is very small.
Furthermore, we would like to analyze
 the behavior in the radiation stage.
In the limit of large $z$,
one has $\Omega_y\rightarrow \Omega_{yi}\simeq (10^{-10},10^{-2})$
and $\Omega_r\rightarrow 1-\Omega_{yi}\simeq1$,
Eqs.(\ref{R}) and (\ref{S}) give an asymptotic behavior
\be
 r\rightarrow3,\ \ \
s\rightarrow\frac{4}{3},\ \ \ q\rightarrow1,
\ee
independent of the initial condition.
The analysis for the 1-loop YMC model
in Ref.\cite{Zhao1} did not include  $\Omega_r$,
so the extrapolation to the radiation stage
could not have been made.

To show the difference of  YMC  from other dark energy models,
Fig.\ref{comp} demonstrates $r(z)$ and $s(z)$  from YMC, LCDM,
quiessence, and kinessence, respectively \cite{Sahni}.
On the left panel,
while YMC gives $r(z)=1\pm 0.02$ around $z\simeq 0$,
quiessence and kinessence give their respective
$r(z)$ decreasing to  $0.5$ and $0.4$.
On the right panel,
for $z \le 4$  in all three cases of YMC model,
 $s(z)$ approaches to $0$, quite close to LCDM model.
On the other hand,
the quiessence
 has a constant $s=0.2$,
and the kinessence has $s(z) > 0.3$ around $z\simeq 0$.
The differences of the YMC
from quiessence and  kinessence are very large.
However, like the other dark energy models,
$s(z)$ in YMC  also has a divergence at $z\sim 13$,
which does not show up in Fig.\ref{comp}.

In the following we discuss the $r-s$ plan and the $r-q$ plan in
the three cases of YMC  model.

\begin{figure}
\centerline{\includegraphics[width=12cm]{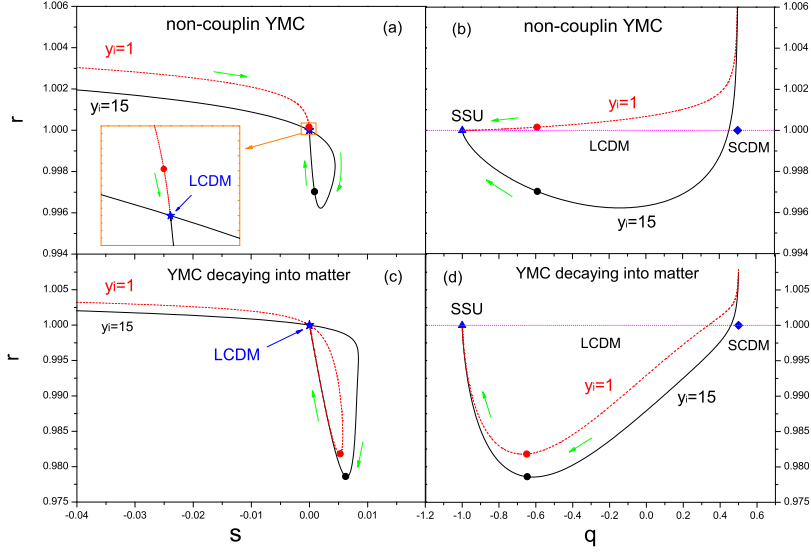}}
 \caption{\label{newfour}\small
 The recent trajectories $(r,s)$ and $(r,q)$
 for the non-coupling case in (a) and (b)
 and for the YMC decaying into  matter in (c) and (d).
 The  round dots are  the current values for
 the $y_i=1$ and for $y_i=15$, respectively.
  The arrows along the curves denote the evolution direction.
 The square dot  at $(\frac{1}{2},1)$ is the fixed point of SCDM,
 and the triangle dot at  $(-1,1)$ is that of the Steady State Universe.
  }
\end{figure}

Fig.\ref{newfour}(a) gives the recent ($z< 0.8$)
 $(r,s)$ trajectory
in the non-coupling case of YMC.
 Two trajectories are presented for two initial values
 $y_i=1$ and $y_i=15$, respectively.
The round dot at $(1.00016, 4.76\times10^{-5})$ is
 the current value for $y_i=1$,
 and that at  $(0.997, 9.07\times10^{-4})$ is
 for $y_i=15$, respectively.
The star at $(r,s)=(1,0)$ is the fixed point of LCDM model,
which currently is still robust against the observations.
Our model predicts an asymptote at $t\rightarrow \infty$
 very close to that of LCDM model.
 The trajectories of  $y_i=15$ and of $y_i=1$
approach to it by different paths.
With the expansion of the universe,
all trajectories for different initial conditions
will reach to the fixed point $(r,s)=(1,0)$ ultimately.
In comparison,
the quiessence  and kinessence models
 give the current values $r<0.5$ and $s\sim 0.5$,
other models have very scattered  typical  values,
such as
the quietessence model \cite{Alam} with  $(r,s)=(0.4,0.3)$,
the Chaplygin gas model \cite{Gorini} with $(r,s)=(1.95,-0.3)$,
the agegraphic model \cite{Cai} with $(r,s)=(-0.2,0.5)$,
which are far away from that of LCDM model \cite{Sahni}.
The holographic model \cite{Zhangjingfei} without interaction
has $(r,s) \simeq (0.94,0.01)$,
but an interaction $b^2=0.1$  gives $(r,s)\simeq (0.75,0.09)$,
deviating  away from LCDM model again.

Fig.\ref{newfour}(b) gives the recent ($z<1.1$) $(r,q)$ trajectory
in the non-coupling case of YMC.
For $y_i=1$, $r$ decreases to unity monotonically,
whereas, for $y_i=15$, $r$ decreases
to a minimum value then increases back to unity.
The two trajectories eventually approach to
the fixed point at $(1, -1)$ of the SSU.
Note that, the pair $(r,q)$ in YMC
does not pass through the fixed point
$(1, \frac{1}{2})$ of the SCDM \cite{Alam}.

Fig.\ref{newfour}(c) shows the  $(r,s)$ trajectory
in the case of YMC decaying into matter,
whose overall profile looks similar to,
but its detail is different from,
that of the non-coupling case in  Fig.\ref{newfour}(a).
The current  values of the Statefinder pair
$(r,s)= (0.981,    5.27\times10^{-3})$ for $y_i=1$,
and $(r,s)= (0.979, 6.21\times10^{-3})$ for $y_i=15$,
respectively.
As $t\rightarrow \infty$,
$(r,s)$ will also approach the fixed point
$(1,0)$ of the LCDM model.

Fig.\ref{newfour}(d) shows the $(r,q)$ trajectory
in the case of YMC decaying into matter,
which is similar to that
of non-coupling case  in Fig.\ref{newfour}(b),
but the case for $y_i=1$ has  $r< 1$
due to the coupling.
As $t\rightarrow \infty$,
$(r,q)$ will also  approach the fixed point $(1,-1)$
of the SSU models.
Moreover,  in this case,
the trajectories of  $(r,q)$ do not pass through
the fixed point$(1, \frac{1}{2})$ of the SCDM either.

The case of  YMC decaying into both matter and radiation.
Their  $r(z)$ and $s(z)$ are plotted in Fig.\ref{sztime}.
Since the YMC-radiation coupling $\Gamma'/H_0 $ is very small,
its modifications to the parameters $r$, $s$, and $q$
are $\le 10^{-3}$.
Therefore, the trajectories $(r,s)$ and $(r,q)$
in this case  are almost overlapped with
Fig.\ref{newfour} (c) and (d), respectively.
To save room, we do not plot the $(r,s)$ and
$(r,q)$ in this case.

\section{Summary}

The Statefinder pair $(r,s)$ is examined for the 2-loop quantum
effective YMC dark energy model. Three cases are presented: the
non-coupling YMC, the YMC decaying into  matter, and the YMC
decaying into both  matter and radiation. It is found  that in all
the three cases the  pair $(r,s)$  is very close to the fixed point
$(1,0)$ of LCDM model for $z<5$, and the deviations are tiny $\delta
r \sim  10^{-2}$ and $\delta s \sim 10^{-2}$. Among the three cases,
$(r,s)$ in the non-coupling case differs  only  by $\sim 1\%$ from
those in the two coupling models, while the two coupling models are
almost the same as each other in all aspects, since the decay rate
of YMC into radiation is very small in the model. In regards to the
diagnosis of Statefinder pair, the YMC model is shown to differ
 drastically  from other dark energy models,
such as
quiessence,  kinessence, quintessence, Chaplygin gas, interacting
holographic, and agegraphic,  etc.
If further cosmological
observations continue to support LCDM model,
they are unlikely to rule out the
YMC dark energy model by using only the pair $(r,s)$.

\

ACKNOWLEDGMENT: Y.Zhang's research work was supported by the CNSF
No.10773009, SRFDP, and CAS.

\

{\bf Appendix }

\

In this appendix we derive the Statefinder  $(r,s)$
for  SCDM with $k=0$, i.e., the Einstein-de Sitter model
containing only  matter.
For the time being, however,
we let the radiation density $\rho_r$ be non-vanishing
and will set $\rho_r=0$ in the final step of calculation.
The Friedman Equations are still given in Eqs.(\ref{fri1})
and (\ref{fri2}), while   the total energy density $\rho=\rho_m+\rho_r$
and the total pressure $p=p_r$.
There is no coupling between  matter and radiation,
so the energy is conserved for each component:
\ba\label{density}
&&\dot{\rho}_m+3H\rho_m=0;\nonumber\\
&&\dot{\rho}_r+3H(\rho_r+p_r)=0,
\ea
which ensure that the total
energy satisfies:
\be \label{total} \dot{\rho}+3H(\rho+p)=0.
 \ee
Taking time derivative of  Eq.(\ref{fri2}) leads to
\be\label{d}
\frac{\stackrel{\dots}{a}}{a}-\frac{\ddot{a}\dot{a}}{a^2}=
-\frac{4\pi G}{3}(\dot{\rho}+3\dot{p}).
\ee
Then with the help of Eqs.(\ref{fri1}),
(\ref{fri2}) and (\ref{total}), one obtains, \be\label{r1}
r\equiv\frac{\stackrel{\dots}{a}}{aH^3}
=1-\frac{3\dot{p}}{2H\rho}. \ee
Applying Eq.(\ref{density}) and taking $\dot{p}=\dot{p}_r=\frac{1}{3}\dot{\rho}_r$
into account,
one obtains $r$ for SCDM model,
\be\label{1}
r
=1+2\Omega_r.
\ee
By Eq. (\ref{fri2}), the deceleration  parameter $q$ for SCDM model
is given by
\be
q  =-\frac{\ddot{a}}{aH^2}
 =\frac{1}{2}(1+\Omega_r).
\ee
Using the definition in Eq.(\ref{RS}), one has
\be
s \equiv \frac{r-1}{3(q-1/2)}
  =   \frac{2\Omega_r}{\frac{3}{2}\Omega_r}=\frac{4}{3},
\ee
which is different from the value
$s=1$ that has been often quoted in literature \cite{Sahni}.
Note that,
since $\Omega_r$ is cancelled,
this result of $s$ is independent of
the value of $\Omega_r$.
Setting $\Omega_r=0$ in Eq.(\ref{1})
yields the Statefinder pair of SCDM
\be
(r,s)=(1,\frac{4}{3}).
\ee

\

\end{document}